\documentclass[journal,12pt,draftclsnofoot,onecolumn,twoside]{IEEEtran}

\usepackage[margin=1.7cm]{geometry} 
\usepackage{ifpdf}
 \ifpdf
     \usepackage[pdftex]{graphicx}
 \else
     \usepackage[dvips]{graphicx}
 \fi

\graphicspath{{figures/}}

\usepackage{cite}
\usepackage[cmex10]{amsmath}
\usepackage{amssymb,amsfonts}
\usepackage{algorithm}
\usepackage{algpseudocode}

\usepackage{mdwmath}
\usepackage{mdwtab}

\usepackage{slashbox}

\newenvironment{frcseries}{\fontfamily{frc}\selectfont}{}

\usepackage[belowskip=-14pt,aboveskip=0pt,small]{caption}

\IEEEaftertitletext{\vspace{-2.5\baselineskip}}

\begin{document}
%
\title{Rate-Interference Tradeoff in OFDM-based Cognitive Radio Networks}

%
%
%
\author{
{Ebrahim Bedeer, 
Octavia A. Dobre, 
Mohamed H. Ahmed, and
Kareem E. Baddour \IEEEauthorrefmark{2}
}\\
\IEEEauthorblockA{Faculty of Engineering and Applied Science, Memorial University,
St. John's, NL, Canada\\
\IEEEauthorrefmark{2} Communications Research Centre, Ottawa, ON, Canada\\
Email: \{e.bedeer, odobre, mhahmed\}@mun.ca, kareem.baddour@crc.ca}
}

\maketitle
\thispagestyle{empty}
\pagestyle{empty}

\begin{abstract}
In cognitive radio (CR) networks, secondary users (SUs) are allowed to opportunistically access the primary users (PUs) spectrum to improve the spectrum utilization; however, this increases the interference levels at the PUs. In this paper, we consider an orthogonal frequency division multiplexing OFDM-based CR network and investigate the tradeoff between increasing the SU transmission rate (hence improving the spectrum utilization) and reducing the interference levels at the PUs. We formulate a new multiobjective optimization (MOOP) problem that jointly maximizes the SU transmission rate and minimizes its transmit power, while imposing interference thresholds to the PUs. Further, we propose an algorithm to strike a balance between the SU transmission rate and the interference levels to the PUs. The proposed algorithm considers the practical scenario of knowing partial channel state information (CSI) of the links between the SU transmitter and the PUs receivers.
Simulation results illustrate the performance of the proposed algorithm and its superiority when compared to the work in the literature.
\end{abstract}


%

\section{Introduction}

The current spectrum underutilization problem is a result of the traditional inefficient spectrum allocation policies rather than the scarcity of the wireless radio spectrum \cite{bedeer2011partial}. To improve the spectrum utilization, the concept of dynamic spectrum access was proposed in recent years  \cite{bedeer2011partial}. Cognitive radio (CR) promoted this concept by allowing secondary (or unlicensed) users (SUs) to opportunistically access the spectrum holes in primary (or licensed) users (PUs) frequency spectrum, subject to constrained degradations to the PUs performance \cite{bedeer2011partial}. Orthogonal frequency division multiplexing (OFDM) is widely recognized as an attractive candidate for the SUs transmission due to
its capabilities in analyzing the spectral activities of PUs \cite{weiss2004spectrum}.

The CR is capable of adapting its transmission to the surrounding environment conditions, with two target objectives \cite{haykin2005cognitive}: 1) improving the spectrum utilization by maximizing the transmission rate of SUs for a given bandwidth and 2) controlling the amount of interference leaked to the PUs receivers due to the SUs transmission. Considering both objectives is a challenging task, as they are conflicting, i.e., increasing the transmission rate of SUs is accompanied by an excessive interference levels to PUs and vice versa. Therefore, there is a tradeoff between the two objectives and it should be carefully investigated in order to have a flexible design that improves the overall performance of the CR networks. In the literature, this design flexibility was not fully exploited, as all the proposed algorithms focused on maximizing the SUs transmission rate, with predefined thresholds for the leaked interference to PUs (i.e., without minimizing the interference to PUs) \cite{bansal2008optimal, zhang2010efficient, attar2008interference, bansal2011adaptive}.

In this paper, we provide a mathematical framework for the rate-interference tradeoff  in the OFDM-based CR networks. This is achieved by formulating a multiobjective optimization (MOOP) problem that jointly maximizes the SU transmission rate and minimizes its transmit power. We additionally set  predefined interference thresholds for each PU as constraints. We consider partial channel-state information (CSI) knowledge on the links between the SU transmitter and the PUs receivers and full CSI knowledge between the SU transmitter and receiver pair.
More specifically, for the SU transmitter and PUs receivers links, we consider the following practical scenarios: 1) knowledge of the path loss and 2) knowledge of the path loss and channel statistics (i.e., the fading distribution and its parameters). For comparison purposes, we additionally consider knowledge of the path loss and full CSI, providing  an upper bound on the SU achievable performance.
We propose a low complexity algorithm to solve the MOOP problem. Simulation results show the performance of the proposed algorithm and illustrate the SU performance degradation due to partial CSI knowledge. Additionally, the results show the advantages  of using the proposed algorithm (in terms of the energy efficiency and the leaked interference to PUs) when compared to other algorithms proposed in the literature.


The remainder of the paper is organized as follows. Section \ref{sec:model} introduces the system model. The MOOP problem is formulated and solved and the proposed algorithm is summarized in Section \ref{sec:opt}.
Simulation results are presented in Section \ref{sec:sim}, while conclusions are drawn in Section \ref{sec:conc}.


\section{System Model} \label{sec:model}

\subsection{System Description}

The available spectrum is divided into $L$ channels that are licensed to $L$ PUs. PUs do not necessarily fully occupy their licensed spectrum temporally and/or spatially; hence, an SU may access such spectrum holes as long as no harmful interference occurs to frequency-adjacent PUs due to adjacent channel interference (ACI) or to other PUs operating in the same frequency band at distant location due to co-channel interference (CCI) \cite{haykin2005cognitive}. Without loss of generality, we assume that the SU decides to use subchannel $m$ of bandwidth $B_m$; this decision can be reached by visiting a database administrated by a third party (e.g., telecomm. authorities), or by optionally sensing the PUs radio spectrum. We assume that the SU accesses subchannel $m$ using OFDM with $N$ subcarriers.


Unlike most of the work done in the literature \cite{bansal2008optimal, zhang2010efficient, attar2008interference}, we assume partial CSI knowledge on the links between the SU transmitter and PUs receivers (this is because estimating the instantaneous channel gains of such links is practically challenging without the PUs cooperation). More specifically, we assume: 1) knowledge of the path loss, which is practically possible especially in applications with stationary nodes. In such a case, the path loss exponent and the node locations can be estimated with high accuracy \cite{salman2012low}; and 2) knowledge of the path loss and channel statistics (i.e., the fading distribution and its parameters), which is a reasonable assumption for certain wireless environments. For example, in non-line-of-sight urban environments, a Rayleigh distribution is usually assumed for the magnitude of the fading channel coefficients \cite{bansal2011adaptive}. The case of full CSI knowledge on the links between the SU transmitter and PUs receivers represents an upper bound on the achievable SU performance and is additionally provided in the numerical results section to characterize the performance loss due to the partial CSI knowledge.
We should note that following the common practice in the literature, we assume that the instantaneous channel gains between the SU transmitter and receiver pair are available through a delay- and error-free feedback channel \cite{bansal2008optimal, attar2008interference, zhang2010efficient, bansal2011adaptive}.

\subsection{Modeling of the CCI and ACI constraints with partial CSI knowledge}


\subsubsection{Case 1---Knowledge of the path loss} The transmit power of the SU on subchannel $m$ should be limited to a certain threshold $P_{th}^{(m)}$ to protect the $m$th distant PU receiver, i.e., $10^{-0.1 \: \textup{PL}(d_{m})}     \sum_{i = 1}^{N} p_i \leq P_{th}^{(m)}$,
where $\textup{PL}(d_m)$ is the distance-based path loss in dB at distance $d_m$ from the SU and $p_i$ is the allocated power per subcarrier $i$, $i = 1, ..., N$.
To reflect the SU transmitter's power amplifier limitations and/or to satisfy regulatory maximum power limits, the SU transmit power should be limited to a certain threshold $P_{th}$, i.e., $\sum_{i = 1}^{N} p_i \leq P_{th}$.
Hence, the constraint on the total transmit power is formulated as $\sum_{i = 1}^{N} p_i \leq \left[P_{th}, \frac{P_{th}^{(m)}}{10^{-0.1 \: \textup{PL}(d_{m})}    } \right]^-$, where $[x,y]^-$ represents $\mathrm{min}(x,y)$.
To simplify the notation and without loss of generality, we assume that $\frac{P_{th}^{(m)}}{10^{-0.1 \: \textup{PL}(d_{m})}    } < P_{th}$. Hence, the CCI constraint is written as
\begin{IEEEeqnarray}{c}
\sum_{i=1}^{N} p_i \leq P_{th}^{(m)} X_{\textup{Case 1}}^{(m)},
\end{IEEEeqnarray}
where $X_{\textup{Case 1}}^{(m)} = \frac{1}{10^{-0.1 \: \textup{PL}(d_{m})}    }$ represents the channel knowledge coefficient from the SU transmitter to the $m$th PU receiver for the case of only knowing the path loss.

The ACI is mainly due to the power spectral leakage of the SU subcarriers to the PUs receivers. This depends on the power allocated to each SU subcarrier and the spectral distance between the SU subcarriers and the PUs receivers \cite{weiss2004spectrum}. The ACI to the $\ell$th PU receiver should be limited to a certain threshold $P_{th}^{(\ell)}$ as $10^{-0.1 \: \textup{PL}(d_{\ell})}     \sum_{i=1}^{N} p_i \: \varpi_i^{(\ell)} \leq P_{th}^{(\ell)}, \quad \ell = 1, ..., L,$
where $\varpi_i^{(\ell)} = T_{s} \: \int_{f_{i,\ell}-\frac{B_\ell}{2}}^{f_{i,\ell}+\frac{B_\ell}{2}} \textup{sinc}^2(T_{s} f) \: df$, $T_s$ is the SU OFDM symbol duration, $f_{i,\ell}$ is the spectral distance between the SU subcarrier $i$ and the $\ell$th PU  frequency band, $B_{\ell}$ is the bandwidth of the $\ell$th PU, and $\textup{sinc}(x) = \frac{\sin(\pi x)}{\pi x}$. The ACI constraint can be further written as
\begin{IEEEeqnarray}{c}
\sum_{i = 1}^{N} p_i \: \varpi_i^{(\ell)} \leq P_{th}^{(\ell)} X_{\textup{Case 1}}^{(\ell)}, \quad \ell = 1, ..., L,
\end{IEEEeqnarray}
where $X_{\textup{Case 1}}^{(\ell)} = \frac{1}{10^{-0.1 \: \textup{PL}(d_{\ell})}    }$ is the channel knowledge coefficient from the SU transmitter to the $\ell$th PU receiver for the case of only knowing the path loss.


\subsubsection{Case 2---Knowledge of the path loss and channel statistics}
The CCI constraint is written as $|\mathcal{H}_{sp}^{(m)} |^2 10^{-0.1 \: \textup{PL}(d_{m})}     \sum_{i = 1}^{N} p_i \leq P_{th}^{(m)}$,
where $\mathcal{H}_{sp}^{(m)}$ is the channel gain to the distant $m$th PU receiver. Since $\mathcal{H}_{sp}^{(m)}$ is not perfectly known at the SU transmitter, the CCI constraint is limited below the threshold $P_{th}^{(m)}$ with at least a probability of $\Psi_{th}^{(m)}$. This is formulated as
\begin{IEEEeqnarray}{c}
\textup{Pr}\left( | \mathcal{H}_{sp}^{(m)} |^2 10^{-0.1 \: \textup{PL}(d_{m})}     \sum_{i = 1}^{N} p_i \leq P_{th}^{(m)} \right) \geq \Psi_{th}^{(m)}. \label{eq:statistical_CCI}
\end{IEEEeqnarray}
A non-line-of-sight propagation environment is assumed; therefore, the channel gain $\mathcal{H}_{sp}^{(m)}$ can be modeled as a zero-mean complex Gaussian random variable, and hence, $| \mathcal{H}_{sp}^{(m)} |^2$ follows an exponential distribution \cite{proakisdigital}. Accordingly, the statistical constraints in (\ref{eq:statistical_CCI}) can be evaluated as
\begin{IEEEeqnarray}{c}
1 - \exp\left(- \frac{\nu^{(m)}}{10^{-0.1 \: \textup{PL}(d_{m})}     \sum_{i = 1}^{N} p_i} P_{th}^{(m)}\right) \geq \Psi_{th}^{(m)}, \label{eq:statistical_CCI_1}
\end{IEEEeqnarray}
where $\frac{1}{\nu^{(m)}}$ is the mean of the exponential distribution.
Equation (\ref{eq:statistical_CCI_1}) can be further simplified as
\begin{IEEEeqnarray}{c}
\sum_{i = 1}^{N} p_i \leq P_{th}^{(m)} X_{\textup{Case 2}}^{(m)},
\end{IEEEeqnarray}
where $X_{\textup{Case 2}}^{(m)} = \frac{\nu^{(m)}}{\left(-\ln(1 - \Psi_{th}^{(m)})\right) 10^{- 0.1 \: \textup{PL}(d_{m})}}$ is the channel knowledge coefficient from the SU transmitter to the $m$th PU receiver for the case of knowing the path loss and the channel statistics. Similarly, the ACI constraint can be written as
\begin{IEEEeqnarray}{c}
\sum_{i = 1}^{N} p_i \: \varpi_i^{(\ell)} \leq P_{th}^{(\ell)} X_{\textup{Case 2}}^{(\ell)}, \quad \ell = 1, ..., L,
\end{IEEEeqnarray}
where $X_{\textup{Case 2}}^{(\ell)} = \frac{\nu^{(\ell)}}{\left(-\ln(1 - \Psi_{th}^{(\ell)})\right) 10^{- 0.1 \: \textup{PL}(d_{\ell})}}$ is the channel knowledge coefficient to the $\ell$th PU for the case of knowing the path loss and the channel statistics.

\section{Optimization Problem and Proposed Algorithm} \label{sec:opt}


Recently, MOOP has attracted researchers' attention due to its flexible and superior performance over single objective optimization approaches \cite{bedeer2013joint}. For most of the MOOP problems, due to the contradiction and incommensurability of the competing objective functions, it is not possible to find a single solution that optimizes all the objective functions simultaneously. In other words, there is no solution that improves one of the objective functions without deteriorating other objectives. However, a set of non-dominated, Pareto optimal solutions exists and it is the decision maker's (the SU in our case) responsibility to choose its preferred optimal solution \cite{miettinen1999nonlinear}.
We solve the MOOP problem by linearly combining the \textit{normalized} competing rate and transmit power objectives  into a single objective function. For that, positive weighting coefficients are used \cite{miettinen1999nonlinear}. These coefficients reflects the SU preferences to the surrounding environment, the wireless application, and/or the target performance.

We formulate a MOOP problem that jointly minimizes the SU transmit power and maximizes its transmission rate, while guaranteeing acceptable levels of CCI and ACI to the existing PUs receivers, as
\begin{IEEEeqnarray}{c}
\underset{p_i}{\textup{min}} \quad \sum_{i = 1}^{N} p_i \qquad \textup{and} \qquad \underset{p_i}{\textup{max}} \quad \sum_{i = 1}^{N} \log_2(1 + p_i \: \frac{|\mathcal{H}_i|^2}{\sigma_n^2 + \mathcal{J}_i}), \nonumber
\end{IEEEeqnarray}
\begin{IEEEeqnarray}{rcl}
\textup{subject to} \qquad \textup{C1}: &{}\quad {}&  \sum_{i = 1}^{N} p_i  \leq  P_{th}^{(m)} X^{(m)}, \nonumber \\
\textup{C2}: &{}\quad {}& \sum_{i = 1}^{N} p_i \varpi_i^{(\ell)} \leq P_{th}^{(\ell)} X^{(\ell)}, \qquad \ell = 1, ..., L, \nonumber \\
\textup{C3}: &{}\quad {}&				 p_i \geq 0, \quad i = 1, ..., N, \label{eq:MOOP_1}
\end{IEEEeqnarray}
where $X^{(m)} \in \{ X_{\textup{Case 1}}^{(m)}, X_{\textup{Case 2}}^{(m)} \}$ and $X^{(\ell)} \in \{ X_{\textup{Case 1}}^{(\ell)}, X_{\textup{Case 2}}^{(\ell)}\}$ represent the channel knowledge coefficients from the SU transmitter to the $m$th and $\ell$th PUs receivers, respectively, $\mathcal{H}_i$ is the channel gain of subcarrier $i, i = 1, ..., N$, between the SU transmitter and receiver pair, $\sigma_n^2$ is the variance of the additive white Gaussian noise (AWGN), and $\mathcal{J}_i$ is the interference from all PUs to the SU subcarrier $i, i = 1, ..., N$ (it depends on the SU receiver windowing function and power spectral density of the PUs \cite{bansal2008optimal, zhang2010efficient,  attar2008interference, bansal2011adaptive}).
The MOOP problem in (\ref{eq:MOOP_1}) can be written as a linear combination of the multiple \emph{normalized} transmit power and rate objectives as
\begin{IEEEeqnarray}{c}
\underset{p_i}{\textup{min}} \quad \alpha \sum_{i = 1}^{N} p_i - (1-\alpha) \sum_{i = 1}^{N} \log_2(1 + \gamma_i p_i), \nonumber \\
\textup{subject to} \qquad \textup{C1---C3}, \label{eq:OP1}
\end{IEEEeqnarray}
where $\gamma_i = \frac{|\mathcal{H}_i|^2}{\sigma_n^2 + \mathcal{J}_i}$ is the channel gain to noise plus interference ratio and $\alpha$ ($0 \leq \alpha \leq 1$) is the weighting coefficient that represents the relative importance of the competing objectives, i.e., higher values of $\alpha$ favor minimizing the transmit power, while lower values of $\alpha$ favor maximizing the rate.
It is worthy to mention that for $\alpha = 0$ the MOOP problem in (\ref{eq:OP1}) reduces to the rate maximization problem in \cite{bansal2008optimal, zhang2010efficient, bansal2011adaptive},  while for $\alpha$ = 1, the optimal solution is zero as the objective is solely to minimize the transmit power.
We assume that the SU chooses the proper value of $\alpha$ depending on the application and/or the surrounding environment. For example, if the transmission rate/time is crucial, then the SU chooses lower values of $\alpha$. On the other hand, if reducing the transmit power/interference to existing PUs (as the sensing process is not fully reliable and/or the channel to the PUs is not perfectly known), protecting the environment, and, hence, the energy efficiency is important, then higher values of $\alpha$ are selected.

\textbf{Proposition 1}: The optimization problem in (\ref{eq:OP1}) is convex and the optimal solution is in the form
\begin{IEEEeqnarray}{c}
p_i^* = \left[\frac{1 - \alpha}{\ln(2) \alpha} - \gamma_i^{-1}\right]^+, \quad i = 1, ..., N, \label{eq:sol_op1_1}
\end{IEEEeqnarray}
if $\sum_{i = 1}^{N} p_i^*  <  P_{th}^{(m)} X^{(m)}
$ and $\sum_{i = 1}^{N} p_i^* \varpi_i^{(\ell)} < P_{th}^{(\ell)} X^{(\ell)}$, $\ell = 1, ..., L,$
where $[x]^+$ represents $\textup{max}(0,x)$; and is  in the form
\begin{IEEEeqnarray}{c}
p_i^* = \left[\frac{1 - \alpha}{\ln(2) \left(\alpha + \lambda_{N+1}\right)} - \gamma_i^{-1}\right]^+, \quad i = 1, ..., N, \label{eq:sol_op1_2}
\end{IEEEeqnarray}
if $\sum_{i = 1}^{N} p_i^*  \geq  P_{th}^{(m)} X^{(m)}$
and $\sum_{i = 1}^{N} p_i^* \varpi_i^{(\ell)} < P_{th}^{(\ell)} X^{(\ell)}, \ell = 1, ..., L,$
where $\lambda_{N+1}$ is calculated to satisfy $\sum_{i = 1}^{N} p_i^* = P_{th}^{(m)} X^{(m)}$; and is in the form
\begin{IEEEeqnarray}{c}
p_i^* = \left[\frac{1 - \alpha}{\ln(2) \left(\alpha + \sum_{\ell = 1}^{L} \lambda_{N+2}^{(\ell)}\varpi_i^{(\ell)}\right)} - \gamma_i^{-1}\right]^+, \nonumber \\ \hfill i = 1, ..., N, \label{eq:sol_op1_3}
\end{IEEEeqnarray}
if $\sum_{i = 1}^{N} p_i^*  <  P_{th}^{(m)} X^{(m)}$
and $\sum_{i = 1}^{N} p_i^* \varpi_i^{(\ell)} \geq P_{th}^{(\ell)} X^{(\ell)}, \ell = 1, ..., L,$
where $\lambda_{N+2}^{(\ell)}$ is calculated to satisfy $\sum_{i = 1}^{N} p_i^* \varpi_i^{(\ell)} = P_{th}^{(\ell)} X^{(\ell)}$, $\ell = 1, ..., L$, and is in the form
\begin{IEEEeqnarray}{c}
p_i^* = \left[\frac{1 - \alpha}{\ln(2) \left(\alpha + \lambda_{N+1} + \sum_{\ell = 1}^{L} \lambda_{N+2}^{(\ell)}\varpi_i^{(\ell)} \right)} - \gamma_i^{-1}\right]^+, \nonumber \\ \hfill i = 1, ..., N, \label{eq:sol_op1_4} \IEEEeqnarraynumspace
\end{IEEEeqnarray}
if $\sum_{i = 1}^{N} p_i^*  \geq  P_{th}^{(m)} X^{(m)}$ and $\sum_{i = 1}^{N} p_i^* \varpi_i^{(\ell)} \geq P_{th}^{(\ell)} X^{(\ell)}, \ell = 1, ..., L,$
where $\lambda_{N+1}$ and $\lambda_{N+2}^{(\ell)}$ are calculated to satisfy $\sum_{i = 1}^{N} p_i^* = P_{th}^{(m)} X^{(m)}$ and $\sum_{i = 1}^{N} p_i^* \varpi_i^{(\ell)} = P_{th}^{(\ell)} X^{(\ell)}$, $\ell = 1, ..., L$, respectively.

\textit{Proof}: See Appendix. \hfill $\blacksquare$


The proposed algorithm can be formally stated as follows:

\floatname{algorithm}{}
\begin{algorithm}
\renewcommand{\thealgorithm}{}
\caption{\textbf{Proposed Algorithm}}
\begin{algorithmic}[1]
\small
\State \textbf{INPUT} $\sigma^2_n$, $\mathcal{H}_{i}$, $\alpha$, $\mathcal{P}_{th}^{(m)}$, $\mathcal{P}_{th}^{(\ell)}$, $X^{(m)}$, and $X^{(\ell)}$, $\ell = 1, ..., L$.
\For{$i$ = 1, ..., $N$}
\State $p_i^*$ is given by (\ref{eq:sol_op1_1}).
\EndFor
\If {$\sum_{i = 1}^{N} p_i^* \geq P_{th}^{(m)} X^{(m)}$ and $\sum_{i = 1}^{N} p_i^* \varpi_i^{(\ell)} < P_{th}^{(\ell)} X^{(\ell)}$}
%
\State $p_i^*$ is given by (\ref{eq:sol_op1_2}).
\State $\lambda_{N+1}$ is calculated to satisfy $\sum_{i = 1}^{N} p_i^* = P_{th}^{(m)} X^{(m)}$.
\ElsIf {$\sum_{i = 1}^{N} p_i^* < P_{th}^{(m)} X^{(m)}$ and $\sum_{i = 1}^{N} p_i^* \varpi_i^{(\ell)} \geq P_{th}^{(\ell)} X^{(\ell)}$}
\State  $p_i^*$ is given by (\ref{eq:sol_op1_3}).
\State $\lambda_{N+2}^{(\ell)}$ is calculated to satisfy $\sum_{i = 1}^{N} p_i^* \varpi_i^{(\ell)} = P_{th}^{(\ell)} X^{(\ell)}$, $\ell = 1, ..., L$.
\Else \:  \textbf{if} {$\sum_{i = 1}^{N} p_i^* \geq P_{th}^{(m)} X^{(m)}$ and $\sum_{i = 1}^{N} p_i^* \varpi_i^{(\ell)} \geq P_{th}^{(\ell)} X^{(\ell)}$} \textbf{then}
\State  $p_i^*$ is given by (\ref{eq:sol_op1_4}).
\State $\lambda_{N+1}$ and $\lambda_{N+2}^{(\ell)}$ are calculated to satisfy $\sum_{i = 1}^{N} p_i^* = P_{th}^{(m)} X^{(m)}$ and $\sum_{i = 1}^{N} p_i^* \varpi_i^{(\ell)} = P_{th}^{(\ell)} X^{(\ell)}$, $\ell = 1, ..., L$, respectively.
\EndIf
\State \textbf{OUTPUT} $p_i^*$, $i$ = 1, ..., $N$.
\end{algorithmic}
\end{algorithm}

The proposed algorithm is briefly explained as follows. Steps 2 to 4 find the optimal solution assuming that both the CCI and ACI constraints are inactive. Based on this assumption, if the CCI constraint is not inactive while the ACI constraints are inactive, the optimal solution is given by steps 5 to 7. Otherwise, if the CCI constraint is inactive and the ACI constraints are not inactive, the optimal solution is given by steps 8 to 10. Finally, if both the CCI and ACI constraints are not inactive the solution is given by steps 11 to 13.

The complexity of the proposed algorithm can be analyzed as follows. The authors in \cite{palomar2005practical} showed that the Lagrange multipliers $\lambda_{N+1}$ and $\lambda_{N+2}^{(\ell)}$, $\ell = 1, ..., L$, that satisfy the CCI and ACI constraints, respectively, can be obtained with linear complexity of the number of subcarrier $N$, i.e., $\mathcal{O}(N)$. Therefore, the computational complexity of the proposed algorithm can be analyzed as follows. Steps 2 to 4 require a complexity of  $\mathcal{O}(N)$; steps 5 to 7, 8 to 10, and 11 to 13 require a complexity of $\mathcal{O}(N^2)$. Thus, the worst case computational complexity  of the proposed algorithm is calculated as $\mathcal{O}(N) + \mathcal{O}(N^2) + \mathcal{O}(N^2) + \mathcal{O}(N^2) = \mathcal{O}(N^2)$. 
\vspace{-10pt}
\section{Numerical Results} \label{sec:sim}


Without loss of generality, we assume that the OFDM SU coexists with a frequency-adjacent PU and a co-channel PU. The SU parameters are: number of subcarriers $N = 128$ and subcarrier spacing $\Delta f = \frac{1.25 \: \rm{MHz}}{N}$. The propagation path loss parameters are: exponent $= 4$, wavelength $= \frac{3 \times 10^8}{900 \times 10^6} = 0.33\:\textup{meters}$, distance between SU transmitter and receiver pair equal to $1$ km, distance to the $\ell$th PU $d_{\ell} = 1.2$ km, distance to the $m$th PU $d_m = 5$ km, and reference distance $d_0 = 100$ m. A Rayleigh fading environment is considered, where the average channel power gains between the SU transmitter and receiver pair $\mathbb{E}\{|\mathcal{H}_i|^2\}$, between the SU transmitter and the receiver of the $\ell$th PU $\mathbb{E}\{|\mathcal{H}_{sp}^{(\ell)}|^2\}$, and between the SU transmitter and the receiver of the $m$th PU $\mathbb{E}\{|\mathcal{H}_{sp}^{(m)}|^2\}$ are set to 0 dB. The PU bandwidth is set to 312.5 kHz. The variance of the AWGN  $\sigma_n^2$ is assumed to be $10^{-15}$ W and the PU signal is assumed to be an elliptically filtered white noise-like process \cite{bansal2008optimal, bansal2011adaptive} of variance $\sigma_n^2$. Representative results are presented in this section, which were obtained through Monte Carlo simulations for $10^{4}$ channel realizations.
Unless otherwise mentioned, the value of the probabilities $\Psi_{th}^{(m)}$ and $\Psi_{th}^{(\ell)}$ is set to 0.9, $P_{th}^{(m)} = 10^{-11}$ W, and $P_{th}^{(\ell`)} = 10^{-11}$ W.
The transmit power and transmission rate objectives are scaled during simulations so
that they are approximately within the same range \cite{miettinen1999nonlinear}. For
convenience, presented numerical results are displayed in the
original scales.

\begin{figure}[!t]
	\centering
		\includegraphics[width=0.5\textwidth]{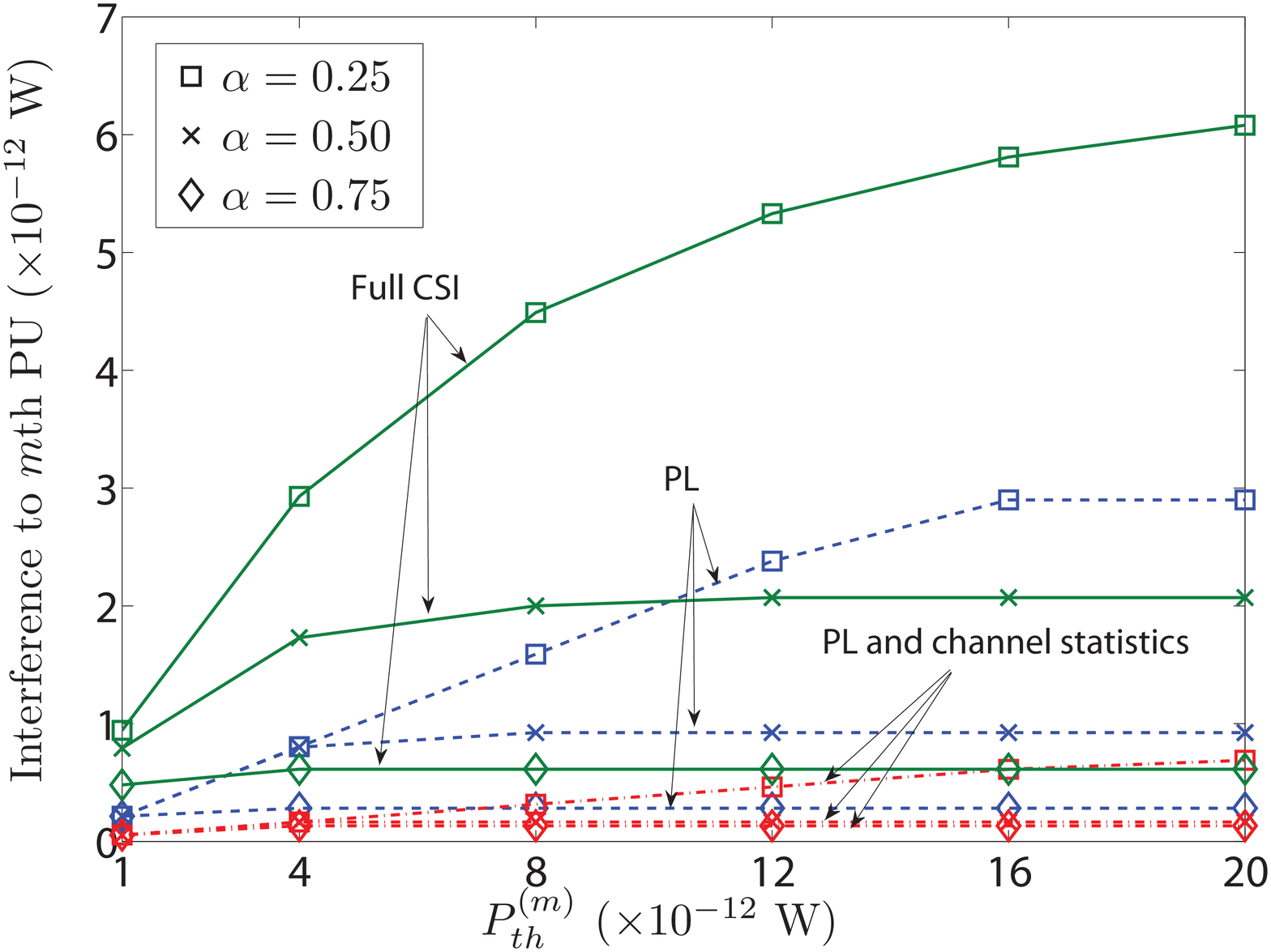}
	\caption{Interference leaked to the $m$th PU as a function of $P_{th}^{(m)}$ for different values of $\alpha$ and for different degree of CSI knowledge.}
	\label{fig:1_min_CCI_PCCI_PCCI}
\end{figure}
\begin{figure}[!t]
	\centering
		\includegraphics[width=0.5\textwidth]{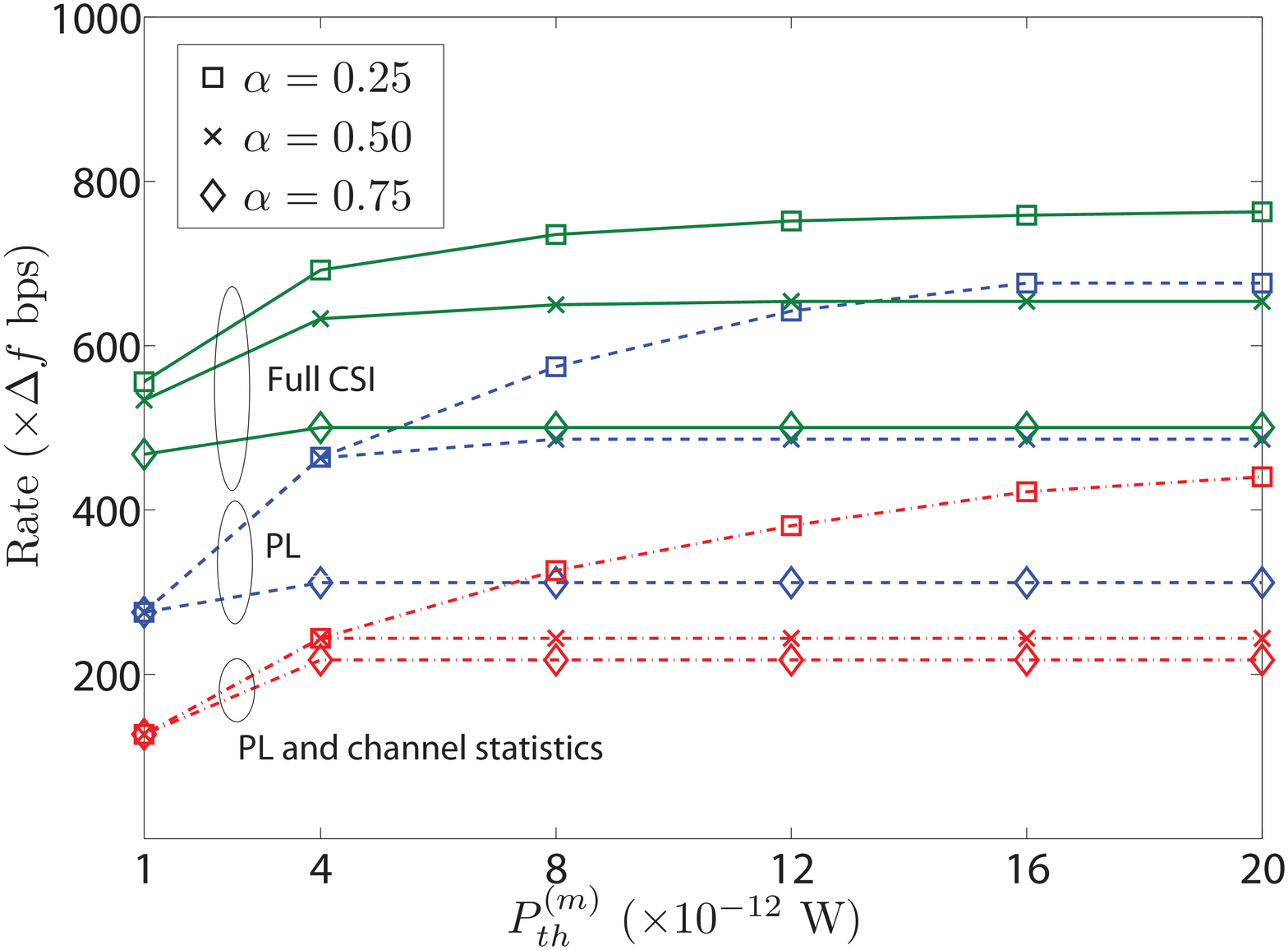}
	\caption{SU rate as a function of $P_{th}^{(m)}$ for different values of $\alpha$ and for different degree of CSI knowledge.}
	\label{fig:1_min_CCI_rate_PCCI}
\end{figure}

Fig. \ref{fig:1_min_CCI_PCCI_PCCI} shows the interference leaked to the $m$th PU receiver as a function of $P_{th}^{(m)}$ for different values of $\alpha$ and for different degrees of CSI knowledge. As can be seen, increasing the value of $\alpha$ reduces the leaked interference to the $m$th PU for all the cases of CSI knowledge. This can be easily explained, as increasing $\alpha$ gives more weight to minimizing the transmit power objective function and less weight to maximizing the transmission rate objective function in (\ref{eq:OP1}). Accordingly, increasing $\alpha$ reduces the CCI to the $m$th PU receiver, but also the SU achievable rate. The interference leaked to the $m$th PU receiver increases linearly with increasing $P_{th}^{(m)}$ for lower values of $P_{th}^{(m)}$ and saturates for higher values of $P_{th}^{(m)}$. This can be explained as follows. For lower values of $P_{th}^{(m)}$, the interference leaked to the $m$th PU receiver is higher than the value of $P_{th}^{(m)}$ and hence, it is limited by the threshold value. On the other hand, for higher values of $P_{th}^{(m)}$, the interference leaked to the $m$th PU receiver is below the threshold value as it is minimized by the proposed algorithm, and hence,  it is kept constant. As expected, knowing the full CSI allows the SU to exploit this knowledge and to transmit with higher power (without violating the interference constraints at the PUs) and higher rate (as it is shown in the discussion of Fig.~\ref{fig:1_min_CCI_rate_PCCI}); this represents an upper bound on the achievable performance. On the other hand, the partial CSI knowledge reduces the transmission opportunities of the SU in order not to violate the interference constraints. It is worthy to mention that the case of knowing only the path loss generates higher interference levels to the existing PUs when compared to the case of knowing the path loss and the channel statistics. This is because the latter case imposes predefined probabilities $\Psi_{th}^{(m)}$ and $\Psi_{th}^{(\ell)}$ on violating the CCI and ACI constraints, respectively, while for the former case the CCI and ACI can be violated uncontrollably. Reducing the values of $\Psi_{th}^{(m)}$ and $\Psi_{th}^{(\ell)}$ produces higher interference levels to the PUs; results are not provided here due to space limitations.

Fig. \ref{fig:1_min_CCI_rate_PCCI} depicts the SU achievable rate as a function of $P_{th}^{(m)}$ for different values of $\alpha$ and for different degrees of CSI knowledge.
Similar to the discussion of Fig. \ref{fig:1_min_CCI_PCCI_PCCI}, the SU achievable rate saturates for higher values of $P_{th}^{(m)}$. This is because the SU transmit power saturates in such a case. As expected, increasing the value of $\alpha$ decreases the SU achievable rate. Further, knowing the full CSI results in higher transmission rate when compared to the partial CSI knowledge.

\begin{figure}[!t]
	\centering
		\includegraphics[width=0.5\textwidth]{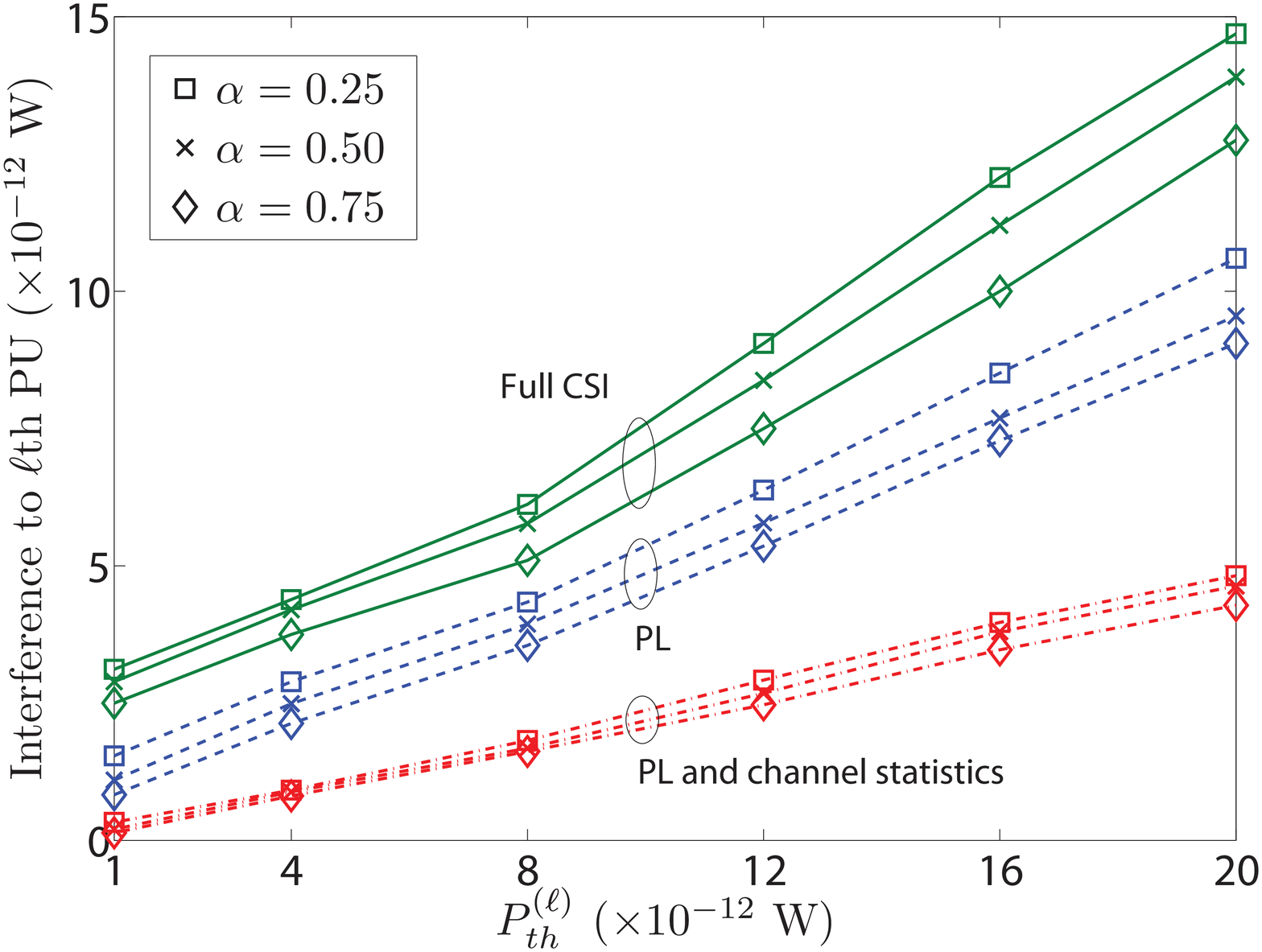}
	\caption{Interference leaked to the $\ell$th PU as a function of $P_{th}^{(\ell)}$ for different values of $\alpha$ and for different degree of CSI knowledge.}
	\label{fig:1_min_CCI_PACI_PACI}
\end{figure}

Fig. \ref{fig:1_min_CCI_PACI_PACI} shows the interference leaked to the $\ell$th PU receiver as a function of $P_{th}^{(\ell)}$ for different values of $\alpha$ and for different degrees of CSI knowledge. As can be seen, increasing the value of $P_{th}^{(\ell)}$ increases the interference leaked to the $\ell$th PU.
As expected, increasing the value of $\alpha$ reduced the interference leaked to the $\ell$th PU receiver and knowing the full CSI enables the SU to transmit higher power and higher transmission rates without violating the interference constraints. The interference leaked to the $\ell$th PU receiver does not saturate for higher values of $P_{th}^{(\ell)}$ as it is not included in the objective function.

\begin{figure}[!t]
	\centering
		\includegraphics[width=0.5\textwidth]{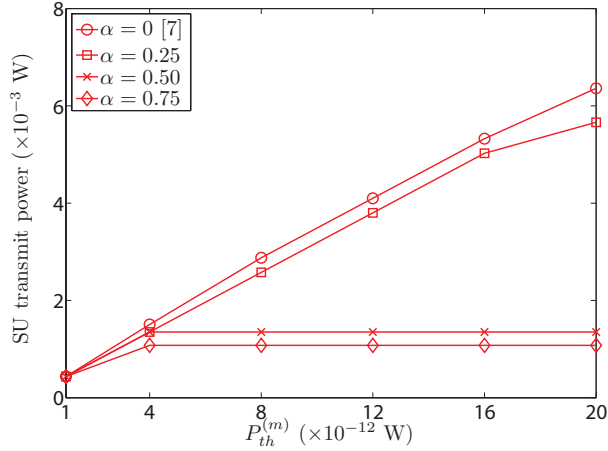}
	\caption{Comparison between the SU transmit power of the proposed algorithm and the algorithm in \cite{bansal2011adaptive}, with the latter corresponding to $\alpha = 0$. }
	\label{fig:1_min_CCI_power_PCCI_comp}
\end{figure}

\begin{figure}[!t]
	\centering
		\includegraphics[width=0.5\textwidth]{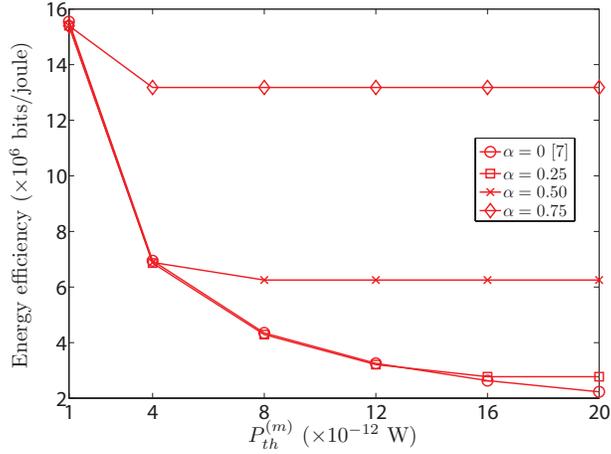}
	\caption{Comparison between the SU energy efficiency of the proposed algorithm and the algorithm in \cite{bansal2011adaptive}, with the latter corresponding to $\alpha = 0$.}
	\label{fig:1_min_CCI_efficiency_PCCI_comp}
\end{figure}

Fig. \ref{fig:1_min_CCI_power_PCCI_comp} compares the SU transmit power of the proposed algorithm with that of the work in \cite{bansal2011adaptive}. It is worthy to mention that the optimization problem in \cite{bansal2011adaptive} can be obtained by setting $\alpha = 0$ in the MOOP problem in (\ref{eq:OP1}). After matching the operating conditions, one can see that the proposed algorithm produces lower SU transmit power; hence, lower interference levels to the $m$th and $\ell$th PU receivers are generated. On the other hand, the work in \cite{bansal2011adaptive} achieves higher SU transmission rate.
However, in Fig. \ref{fig:1_min_CCI_efficiency_PCCI_comp}, the energy efficiency (in bits/joule) of the work in \cite{bansal2011adaptive} and that of the proposed work are compared for the same operating conditions. As can be noticed, the proposed algorithm is more energy efficient when compared to the work in \cite{bansal2011adaptive} with even less complexity (the complexity of the algorithm in \cite{bansal2011adaptive} is $\mathcal{O}(N^3)$).
The energy efficiency saturates for the same range over which the SU transmit power saturates, as seen in Fig. \ref{fig:1_min_CCI_power_PCCI_comp}.

\vspace{-10pt}
\section{Conclusions} \label{sec:conc}
In this paper, we considered an OFDM-based CR network and adopted a multiobjective optimization approach to investigate the tradeoff between increasing the SU transmission rate and reducing the SU transmit power (hence the interference to the PUs). This formulation is considered as a generalization of the work in the literature that focused only on maximizing the SU transmission rate. Simulation results showed the flexibility of the proposed algorithm, which can provide different SU rates and interference levels to the PUs. Further, results showed that the obtained solution is more energy efficient when compared with that of other works in the literature, at the cost of no additional complexity.
In future work, we plan to extended the MOOP approach to the case of multiple SUs.
\vspace{-10pt}



\section*{Appendix: Proof of Proposition 1}

The MOOP problem in (\ref{eq:OP1}) is convex and it can be solved by applying the Karush-Khun-Tucker (KKT) conditions (i.e., transforming the inequalities constraints to equality constraints by adding non-negative slack variables) \cite{Boyd2004convex}. The Lagrangian function $\mathcal{L}(\mathbf{p},\mathbf{y},\boldsymbol \lambda)$ is expressed as
\begin{IEEEeqnarray}{rcl}
\mathcal{L}(\mathbf{p},\mathbf{y},\boldsymbol \lambda) &{} = {}& \alpha \sum_{i = 1}^{N} p_i - (1-\alpha) \sum_{i = 1}^{N} \log_2(1 + \gamma_i p_i) \nonumber
\end{IEEEeqnarray}
\begin{IEEEeqnarray}{rcl}
&{} {}& + \lambda_i \left[-p_i + y_i^2\right] + \lambda_{N+1} \left[\sum_{i = 1}^{N} p_i - P_{th}^{(m)} X^{(m)} + y_{N+1}^2 \right] \nonumber \\
&{} {}& + \sum_{\ell = 1}^{L} \lambda_{N+2}^{(\ell)} \left[\sum_{i = 1}^{N} p_i \varpi_i^{(\ell)} -  P_{th}^{(\ell)} X^{(\ell)} + (y_{N+2}^{(\ell)})^2 \right],
\end{IEEEeqnarray}
where $\mathbf{y} = \left[y_1^2, ..., y_{N+1}^2, (y_{N+2}^{(\ell)})^2 \right]^T$ and $\boldsymbol \lambda = \left[\lambda_1, ..., \lambda_{N+1}, \lambda_{N+2}^{(\ell)} \right]^T$, $\ell = 1, ..., L$, are the vectors of the slack variables and Lagrange multipliers of length  $N + L + 1$. The optimal solution is found when $\nabla \mathcal{L}(\mathbf{p},\mathbf{y},\boldsymbol \lambda) = 0$ as
\begin{subequations}
\begin{IEEEeqnarray}{rcl}
\frac{\partial \mathcal{L}}{\partial p_i} &{}={}& \alpha - \frac{(1-\alpha)}{\ln(2) (p_i + \gamma_i^{-1})} - \lambda_i + \lambda_{N+1} \nonumber \\ & & \hfill + \sum_{\ell = 1}^{L} \lambda_{N+2}^{(\ell)} \varpi_i^{(\ell)} = 0, \label{eq:OP_1_first} \\
\frac{\partial \mathcal{L}}{\partial \lambda_i} &{}={}& - p_i + y_i^2 = 0, \\
\frac{\partial \mathcal{L}}{\partial \lambda_{N+1}} &{}={}& \sum_{i = 1}^{N} p_i - P_{th}^{(m)} X^{(m)} + y_{N+1}^2 = 0, \\
\frac{\partial \mathcal{L}}{\partial \lambda_{N+2}^{(\ell)}} &{}={}& \sum_{i = 1}^{N} p_i \varpi_i^{(\ell)} -  P_{th}^{(\ell)} X^{(\ell)} + (y_{N+2}^{(\ell)})^2 = 0, \\
\frac{\partial \mathcal{L}}{\partial y_i} &{}={}& 2 \lambda_i y_i = 0, \label{eq:OP_1_5}\\
\frac{\partial \mathcal{L}}{\partial y_{N+1}} &{}={}& 2 \lambda_{N+1} y_{N+1} = 0, \label{eq:OP_1_6}\\
\frac{\partial \mathcal{L}}{\partial y_{N+2}} &{}={}& 2 \lambda_{N+2}^{(\ell)} y_{N+2}^{(\ell)} = 0. \label{eq:OP_1_last}
\end{IEEEeqnarray}
\end{subequations}
It can be seen that (\ref{eq:OP_1_first})--(\ref{eq:OP_1_last}) represent $3 N + 2 L + 2$ equations in the $3 N + 2 L + 2$ unknown components of the vectors $\mathbf{p}, \mathbf{y}$, and $\boldsymbol \lambda$. Equation (\ref{eq:OP_1_5}) implies that either $\lambda_i = 0$ or $y_i = 0$, (\ref{eq:OP_1_6}) implies that either $\lambda_{N+1} = 0$ or $y_{N+1} = 0$, and (\ref{eq:OP_1_last}) implies that either $\lambda_{N+2}^{(\ell)} = 0$ or $y_{N+2}^{(\ell)} = 0$, $\ell = 1, ..., L$. Hence, eight possible cases exist, as follows:

---\textit{Case 1}: Setting $\lambda_i = 0$ (i.e., $p_i > 0$), $\lambda_{N+1} = 0$ (i.e., $\sum_{i = 1}^{N} p_i < P_{th}^{(m)} X^{(m)}$), and $\lambda_{N+2}^{(\ell)} = 0$ (i.e., $\sum_{i = 1}^{N} p_i \varpi_i^{(\ell)} < P_{th}^{(\ell)} X^{(\ell)}$) results in the optimal solution on the form
\begin{IEEEeqnarray}{c}
p_i^* = \left[\frac{1 - \alpha}{\ln(2) \alpha} - \gamma_i^{-1}\right]^+, \quad i = 1, ..., N.
\end{IEEEeqnarray}

---\textit{Case 2}: Setting $\lambda_i = 0$ (i.e., $p_i > 0$), $y_{N+1} = 0$ (i.e., $\sum_{i = 1}^{N} p_i = P_{th}^{(m)} X^{(m)}$), and $\lambda_{N+2}^{(\ell)} = 0$ (i.e., $\sum_{i = 1}^{N} p_i \varpi_i^{(\ell)} < P_{th}^{(\ell)} X^{(\ell)}$) results in the optimal solution on the form
\begin{IEEEeqnarray}{c}
p_i^* = \left[\frac{1 - \alpha}{\ln(2) \left(\alpha + \lambda_{N+1}\right)} - \gamma_i^{-1}\right]^+, \quad i = 1, ..., N,
\end{IEEEeqnarray}
where $\lambda_{N+1}$ is calculated to satisfy $\sum_{i = 1}^{N} p_i^* = P_{th}^{(m)} X^{(m)}$.

---\textit{Case 3}: Setting $\lambda_i = 0$ (i.e., $p_i > 0$), $\lambda_{N+1} = 0$ (i.e., $\sum_{i = 1}^{N} p_i < P_{th}^{(m)} X^{(m)}$), and $y_{N+2}^{(\ell)} = 0$ (i.e., $\sum_{i = 1}^{N} p_i \varpi_i^{(\ell)} = P_{th}^{(\ell)} X^{(\ell)}$) results in the optimal solution on the form
\begin{IEEEeqnarray}{c}
p_i^* = \left[\frac{1 - \alpha}{\ln(2) \left(\alpha + \sum_{\ell = 1}^{L} \lambda_{N+2}^{(\ell)}\varpi_i^{(\ell)}\right)} - \gamma_i^{-1}\right]^+,
\nonumber \\ \hfill i = 1, ..., N,
\end{IEEEeqnarray}
where $\lambda_{N+2}^{(\ell)}$ are calculated to satisfy $\sum_{i = 1}^{N} p_i \varpi_i^{(\ell)} = P_{th}^{(\ell)} X^{(\ell)}$, $\ell = 1, ..., L$.

---\textit{Case 4}: Setting $\lambda_i = 0$ (i.e., $p_i > 0$), $y_{N+1} = 0$ (i.e., $\sum_{i = 1}^{N} p_i = P_{th}^{(m)} X^{(m)}$), and $y_{N+2}^{(\ell)} = 0$ (i.e., $\sum_{i = 1}^{N} p_i \varpi_i^{(\ell)} = P_{th}^{(\ell)} X^{(\ell)}$) results in the optimal solution on the form
\begin{IEEEeqnarray}{c}
p_i^* = \left[\frac{1 - \alpha}{\ln(2) \left(\alpha + \lambda_{N+1} + \sum_{\ell = 1}^{L} \lambda_{N+2}^{(\ell)}\varpi_i^{(\ell)} \right)} - \gamma_i^{-1}\right]^+, \nonumber \\ \hfill i = 1, ..., N, \IEEEeqnarraynumspace
\end{IEEEeqnarray}
where $\lambda_{N+1}$ and $\lambda_{N+2}^{(\ell)}$ are calculated to satisfy $\sum_{i = 1}^{N} p_i^* = P_{th}^{(m)} X^{(m)}$ and $\sum_{i = 1}^{N} p_i \varpi_i^{(\ell)} = P_{th}^{(\ell)} X^{(\ell)}$, respectively.

---\textit{Case 5}: Setting $y_i = 0$ (i.e., $p_i = 0$), $\lambda_{N+1} = 0$ (i.e., $\sum_{i = 1}^{N} p_i < P_{th}^{(m)} X^{(m)}$), and $\lambda_{N+2}^{(\ell)} = 0$ (i.e., $\sum_{i = 1}^{N} p_i \varpi_i^{(\ell)} < P_{th}^{(\ell)} X^{(\ell)}$) results in the optimal solution $p_i^* = 0$.

---\textit{Case 6}: Setting $y_i = 0$ (i.e., $p_i = 0$), $y_{N+1} = 0$ (i.e., $\sum_{i = 1}^{N} p_i = P_{th}^{(m)} X^{(m)}$), and $\lambda_{N+2}^{(\ell)} = 0$ (i.e., $\sum_{i = 1}^{N} p_i \varpi_i^{(\ell)} < P_{th}^{(\ell)} X^{(\ell)}$) is invalid as it implies that $p_i^* = 0$ which violates $\sum_{i = 1}^{N} p_i^* = P_{th}^{(m)} X^{(m)}$, $P_{th}^{(m)} \neq 0$.

---\textit{Case 7}: Setting $y_i = 0$ (i.e., $p_i = 0$), $\lambda_{N+1} = 0$ (i.e., $\sum_{i = 1}^{N} p_i < P_{th}^{(m)} X^{(m)}$), and $y_{N+2}^{(\ell)} = 0$ (i.e., $\sum_{i = 1}^{N} p_i \varpi_i^{(\ell)} = P_{th}^{(\ell)} X^{(\ell)}$) is invalid as it implies that $p_i^* = 0$ which violates $\sum_{i = 1}^{N} p_i \varpi_i^{(\ell)} = P_{th}^{(\ell)} X^{(\ell)}$, $P_{th}^{(\ell)} \neq 0$, $\ell = 1, ..., L$.

---\textit{Case 8}: Setting $y_i = 0$ (i.e., $p_i = 0$), $y_{N+1} = 0$ (i.e., $\sum_{i = 1}^{N} p_i = P_{th}^{(m)} X^{(m)}$), and $y_{N+2}^{(\ell)} = 0$ (i.e., $\sum_{i = 1}^{N} p_i \varpi_i^{(\ell)} = P_{th}^{(\ell)} X^{(\ell)}$) is invalid as it implies that $p_i^* = 0$ which violates $\sum_{i = 1}^{N} p_i^* = P_{th}^{(m)} X^{(m)}$, $P_{th}^{(m)} \neq 0$ and $\sum_{i = 1}^{N} p_i \varpi_i^{(\ell)} = P_{th}^{(\ell)} X^{(\ell)}$, $P_{th}^{(\ell)} \neq 0$, $\ell = 1, ..., L$.

The solution $p_i^*$ satisfies the KKT conditions \cite{Boyd2004convex}, and, hence, it is an optimal solution (the proof is not provided due to space limitations).
\vspace*{-20pt}


\begin{thebibliography}{10}
	\providecommand{\url}[1]{#1}
	\csname url@samestyle\endcsname
	\providecommand{\newblock}{\relax}
	\providecommand{\bibinfo}[2]{#2}
	\providecommand{\BIBentrySTDinterwordspacing}{\spaceskip=0pt\relax}
	\providecommand{\BIBentryALTinterwordstretchfactor}{4}
	\providecommand{\BIBentryALTinterwordspacing}{\spaceskip=\fontdimen2\font plus
		\BIBentryALTinterwordstretchfactor\fontdimen3\font minus
		\fontdimen4\font\relax}
	\providecommand{\BIBforeignlanguage}[2]{{%
			\expandafter\ifx\csname l@#1\endcsname\relax
			\typeout{** WARNING: IEEEtran.bst: No hyphenation pattern has been}%
			\typeout{** loaded for the language `#1'. Using the pattern for}%
			\typeout{** the default language instead.}%
			\else
			\language=\csname l@#1\endcsname
			\fi
			#2}}
	\providecommand{\BIBdecl}{\relax}
	\BIBdecl
	
	\bibitem{bedeer2011partial}
	E.~Bedeer, M.~Marey, O.~Dobre, and K.~Baddour, ``On partially overlapping
	coexistence for dynamic spectrum access in cognitive radio,'' in \emph{Proc.
		{IEEE} CAMAD}, Jun. 2011, pp. 143--147.
	
	\bibitem{weiss2004spectrum}
	T.~A. Weiss and F.~K. Jondral, ``Spectrum pooling: an innovative strategy for
	the enhancement of spectrum efficiency,'' \emph{{IEEE} Commun. Mag.},
	vol.~42, no.~3, pp. S8--14, Mar. 2004.
	
	\bibitem{haykin2005cognitive}
	S.~Haykin, ``Cognitive radio: brain-empowered wireless communications,''
	\emph{{IEEE} J. Sel. Areas Commun.}, vol.~23, no.~2, pp. 201--220, Feb. 2005.
	
	\bibitem{bansal2008optimal}
	G.~Bansal, M.~Hossain, and V.~Bhargava, ``Optimal and suboptimal power
	allocation schemes for {OFDM}-based cognitive radio systems,'' \emph{{IEEE}
		Trans. Wireless Commun.}, vol.~7, no.~11, pp. 4710--4718, Nov. 2008.
	
	\bibitem{zhang2010efficient}
	Y.~Zhang and C.~Leung, ``An efficient power-loading scheme for {OFDM}-based
	cognitive radio systems,'' \emph{{IEEE} Trans. Veh. Technol.}, vol.~59,
	no.~4, pp. 1858--1864, May 2010.
	
	\bibitem{attar2008interference}
	A.~Attar, O.~Holland, M.~Nakhai, and A.~Aghvami, ``Interference-limited
	resource allocation for cognitive radio in orthogonal frequency-division
	multiplexing networks,'' \emph{IET Commun.}, vol.~2, no.~6, pp. 806--814,
	Jul. 2008.
	
	\bibitem{bansal2011adaptive}
	G.~Bansal, M.~Hossain, and V.~Bhargava, ``Adaptive power loading for
	{OFDM}-based cognitive radio systems with statistical interference
	constraint,'' \emph{{IEEE} Trans. Wireless Commun.}, no.~99, pp. 1--6, Sep.
	2011.
	
	\bibitem{salman2012low}
	N.~Salman, A.~Kemp, and M.~Ghogho, ``Low complexity joint estimation of
	location and path-loss exponent,'' \emph{IEEE Wireless Commun. Lett.},
	vol.~1, no.~4, pp. 364--367, Aug. 2012.
	
	\bibitem{proakisdigital}
	J.~Proakis, \emph{{Digital Communications}}.\hskip 1em plus 0.5em minus
	0.4em\relax McGraw-Hill, New York NY, 2001.
	
	\bibitem{bedeer2013joint}
	E.~Bedeer, O.~Dobre, M.~H. Ahmed, and K.~E. Baddour, ``Joint optimization of
	bit and power loading for multicarrier systems,'' \emph{IEEE Wireless Commun.
		Lett.}, vol.~2, no.~4, pp. 447--450, Aug. 2013.
	
	\bibitem{miettinen1999nonlinear}
	K.~Miettinen, \emph{Nonlinear Multiobjective Optimization}.\hskip 1em plus
	0.5em minus 0.4em\relax Springer, 1999.
	
	\bibitem{palomar2005practical}
	D.~P. Palomar and J.~R. Fonollosa, ``Practical algorithms for a family of
	waterfilling solutions,'' \emph{{IEEE} Trans. Signal Process.}, vol.~53,
	no.~2, pp. 686--695, Feb. 2005.
	
	\bibitem{Boyd2004convex}
	S.~Boyd and L.~Vandenberghe, \emph{Convex Optimization}.\hskip 1em plus 0.5em
	minus 0.4em\relax Cambridge University Press, 2004.
	
\end{thebibliography}
\end{document}